\documentclass[reprint,aps,prb,superscriptaddress]{revtex4-2}
\usepackage[pdftex, colorlinks=true, linkcolor=blue, citecolor=blue, urlcolor=blue]{hyperref}
\usepackage{graphicx}
\usepackage{aurical}
\usepackage{amsmath, amssymb}
\usepackage{physics}
\usepackage[T1]{fontenc}
\usepackage[version=3]{mhchem}
\usepackage{siunitx}
\usepackage{braket}
\graphicspath{{./figures/}}
\DeclareGraphicsExtensions{.png,.pdf}

\usepackage{color}

\newcommand{\una}{Universit\'{e} de Nouakchott, Facult\'{e} des Sciences et Techniques, D\'{e}partement de Physique, Avenue du Roi Fai\c{c}al, 2373, Nouakchott, Mauritania}

\begin{document}

\title{Unconventional alternating out-of-plane spin polarization in the coplanar kagome antiferromagnet}

\author{O. Ly}
\email{ousmanebouneoumar@gmail.com}
\affiliation{\una}
\author{S. Hayami}
\email{hayami@sci.hokudai.ac.jp}
\affiliation{Graduate School of Science, Hokkaido University, Sapporo 060-0810, Japan.}

\begin{abstract}
The emergence of spin-polarized currents in nonrelativistic platforms continues to attract significant interest in spintronics. Here we demonstrate that a noncollinear kagome antiferromagnet can generate an alternating out-of-plane spin polarization originating from the spin chirality of the magnetic unit cell, in the absence of relativistic spin--orbit coupling. Under spatial confinement, the system develops distinct real-space spin separation patterns whose structure is governed by the symmetry of the lattice termination. In particular, breaking the transverse mirror symmetry of the ribbon produces an altermagnetic-like spin splitting in the band structure. Furthermore, we uncover a spin--edge locking mechanism in which propagating edge states acquire an unconventional spin polarization. These results highlight how magnetic symmetry and confinement can generate spin-polarized transport in coplanar antiferromagnets without relying on relativistic interactions.
\end{abstract}

\maketitle

\section{Introduction}
The electronic spin degree of freedom underlies a wide range of phenomena that continue to drive the development of spintronics \cite{Zutic2004,Wolf2001,Hirohata2020}. Early advances in this field largely relied on relativistic spin--orbit coupling (SOC) \cite{Manchon2015}, which is responsible for fundamental effects such as the anomalous Hall effect and the spin Hall effect (SHE) \cite{Nagaosa2010,Sinova2015}. More recently, similar transport phenomena have been predicted and observed in a class of noncollinear antiferromagnets, where a nonvanishing Berry curvature leads to anomalous \cite{Chen2014,Nakatsuji2015,Nayak2016} and spin Hall \cite{Zelezny2017, Zhang2018} responses.
In these systems, nontrivial magnetic textures can induce effective spin--momentum locking and give rise to a variety of topological transport effects \cite{Baltz2018,Bonbien2022}.

While the real-space manifestation of the conventional spin Hall effect—namely the accumulation of opposite spin polarizations at the edges of a conductor—has been investigated in paramagnetic systems \cite{Nikolic2006,Sinova2015}, its counterpart in non-collinear antiferromagnets remains largely unexplored. 
Although the emergence of unconventional out-of-plane spin currents in the coplanar setup has been studied \cite{Zhang2018}, {it remains unclear how such momentum-space spin responses manifest themselves in real space, particularly under spatial confinement.} 

In this work, we investigate the real-space behavior of out-of-plane spin polarization in a kagome antiferromagnet with a coplanar $120^\circ$ magnetic configuration. We perform numerical simulations of the electronic spin polarization in confined ribbon geometries. We find that spin Hall currents in these topological magnets exhibit unconventional behavior both in momentum and in real space. Under appropriate confinement symmetries, the spin polarization develops alternating spatial patterns across the ribbon, leading to a nontrivial spin separation within the system.

A notable feature of the symmetric geometry is that a single propagating mode exhibits spatial spin separation across the ribbon. In contrast to the conventional quantum spin Hall effect, where two spin-polarized channels propagate along opposite edges of the sample \cite{Kane2005,Bernevig2006}, here a single transport channel develops a position-dependent spin polarization that changes sign across the wire. This behavior reflects the spin–sublattice entanglement induced by the non-collinear magnetic texture.
Furthermore, when the ribbon termination becomes asymmetric in the transverse direction, the system develops an altermagnetic-like spin splitting. This provides an example of how altermagnetic splitting can emerge from real-space confinement.

Our results reveal an overlooked spin separation mechanism in non-collinear kagome antiferromagnets and highlight the role of magnetic symmetry and confinement in shaping spin transport. 
More broadly, we establish topological antiferromagnetic systems as promising platforms for realizing unconventional non-relativistic spin phenomena relevant for future spintronic applications.


\section{Model and methodology}
To study spin transport in the kagome antiferromagnetic system, we start from a tight-binding Hamiltonian
\begin{equation}
H = -\gamma \sum_{\langle i,j\rangle} c_i^\dagger c_j 
+ J \sum_i c_i^\dagger (\mathbf m_i \cdot \boldsymbol\sigma) c_i ,
\end{equation}
where $\mathbf m_i$ denotes the local magnetic moments coupled to the itinerant electrons via an $s$--$d$ exchange interaction with coupling strength $J$ { and $\boldsymbol\sigma$ stands for the vector of Pauli matrices in spin-space}. The operators $c_i$ and $c_i^\dagger$ are the annihilation and creation operators at lattice site $i$, while $\gamma$ is the nearest-neighbor hopping amplitude. { The s-d exchange constant is fixed at $J=3\gamma$ throughout the manuscript. }

\begin{figure}
    \includegraphics[width=0.5\textwidth]{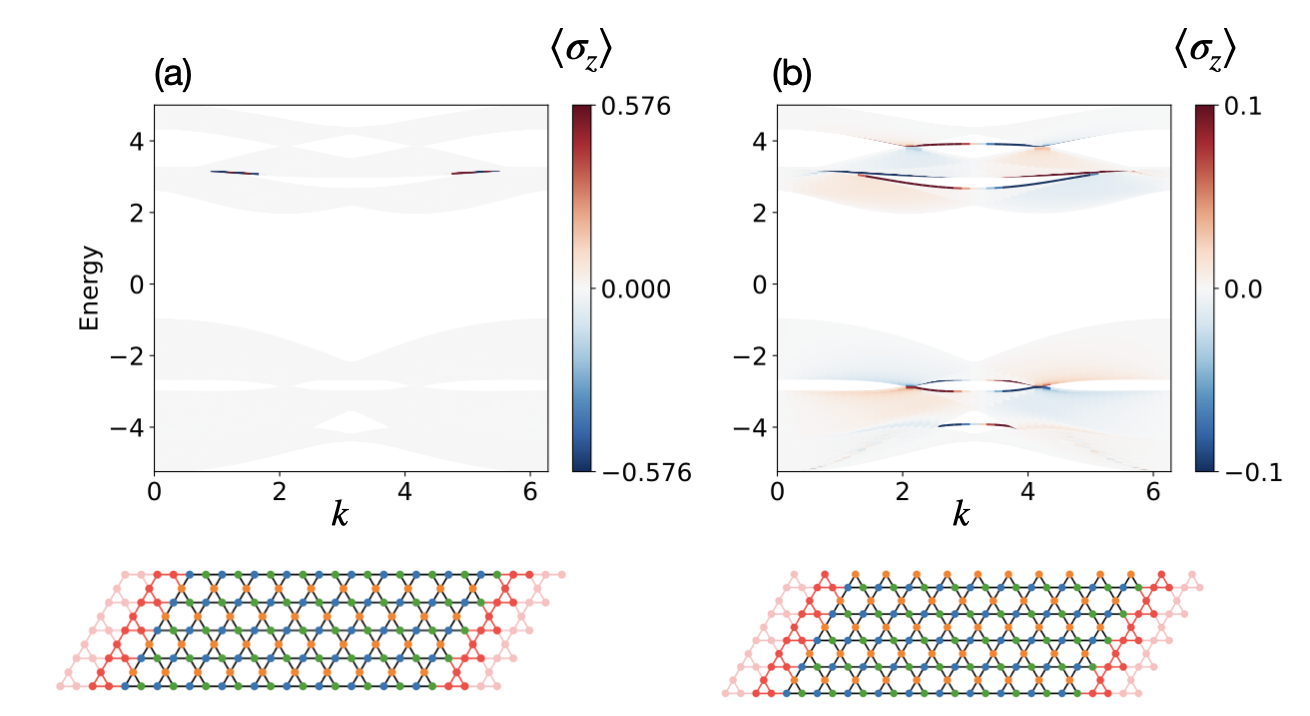}
    \caption{A sketch of the investigated ribbon with symmetric (a) and antisymmetric (b) edge terminations. The corresponding ribbon band structures are shown.}
    \label{fig1}
\end{figure}

For the coplanar $120^\circ$ magnetic configuration, the local moments on the three kagome sublattices can be chosen as $\boldsymbol{m}_i=(\cos\phi_i, \sin\phi_i, 0)$, such that each two of them make an angle of $120^{\circ}$. 
{The three magnetic moments lie in the $xy$ plane, while the $z$ axis is taken to be perpendicular to the magnetic plane.}
Because the three magnetic moments point along different directions, one assumes that no global spin quantization axis is preserved and the electronic eigenstates are intrinsically spin mixed.

The resulting Bloch Hamiltonian therefore cannot be decomposed into independent spin sectors, and the eigenstates correspond to magnetic Bloch spinors defined in the combined sublattice-spin space.

In a ribbon geometry, the confined wave functions can display nontrivial spin-resolved spatial modulations. Even a single propagating mode corresponds to an eigenvector of the Bloch Hamiltonian whose projection onto a fixed $(\uparrow,\downarrow)$ basis generally involves position-dependent interference between spin and sublattice components. As a consequence, the resulting real-space spin polarization is difficult to analyze analytically.
Therefore, we resort to a numerical analysis.
 
To this end, we use the quantum transport framwork \texttt{Kwant} \cite{kwant}. 
Our interest will be in the calculation of local out-of-plane spin polarization generated by an incoming 
{scattering state of a given} lead ($l$) at transport energy $E$. 
This is obtained from the corresponding scattering wave functions $\psi_{i}(E)$ { evaluated at site $i$} as
\begin{equation}
\langle \sigma^i_z \rangle = \psi_{i}^{\dagger}(E)\, \sigma_z\, \psi_{i}(E),
\end{equation}
where a sum over all propagating modes at transport energy $E$ is assumed.
{
To further connect the local spin polarization to spin transport, one may also introduce the {real-space resolved} spin-current associated with the $\alpha$-spin component flowing along the bond $i \to j$ as
{
\begin{equation}
J^{\alpha}_{ij} = 2 \Im[\psi_{i}^{\dagger}\sigma_\alpha H_{ij} \psi_{j}],
\end{equation}
where $H_{ij}$ is the hoping matrix between $i$ and $j$. { A summation over all propagating modes is also assumed.}} 
In particular, the out-of-plane spin transport is characterized by $J^z_{ij}$. Although spin current is not generally conserved in the presence of spin-dependent hopping, this quantity provides a useful measure of spin-resolved transport in the present noncollinear magnetic system.}

\begin{figure*}
    \includegraphics[width=\textwidth]{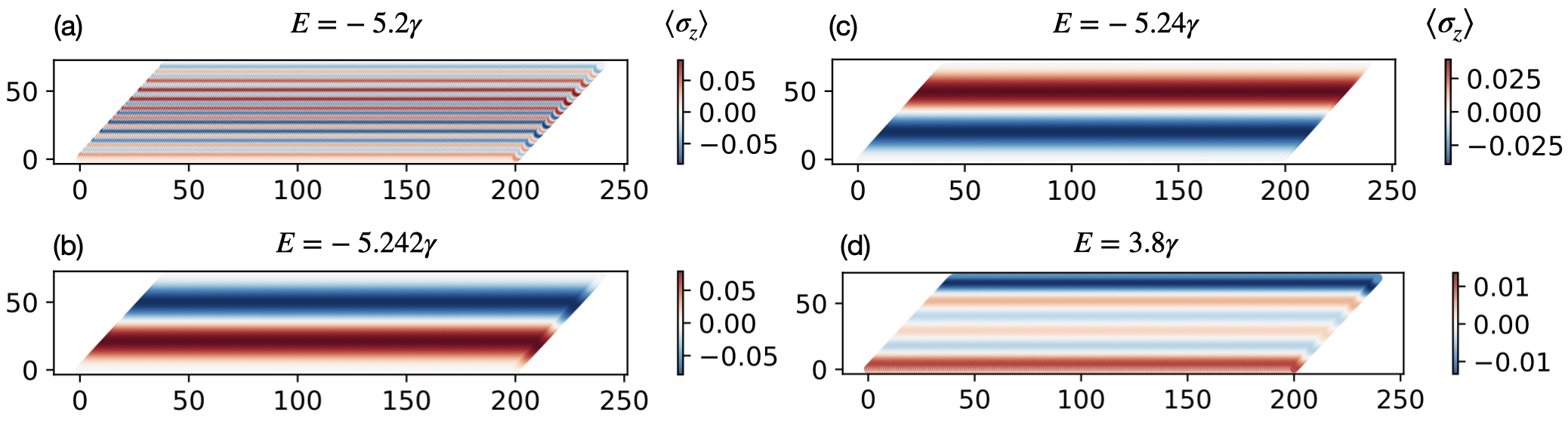}
    \caption{Real-space spin densities in the ribbon for different transport energies ({ E}), corresponding to the propagation of $20$ modes (a) and $2$ modes (b) for the reflection symmetric ribbon. In the right panel the spin polarization in the asymmetric ribbon is displayed for two different energies corresponding to $2$ (c) and $3$ (d) propagating modes. { The ribbon geometry has width $W=80$ and length $L=200$ unit cells.}}
    \label{fig2}
\end{figure*}

\section{Results}
Conventionally, the emergence of {transverse} spin currents (SHE) in magnetic systems requires relativistic SOC. However, the essential ingredient underlying this phenomenon is simply the breaking of spin rotational symmetry as argued in ref. \cite{Zhang2018}. In the presence of a coplanar magnetic order, the circulation of electronic wave functions within the unit cell acquires a nontrivial topological phase induced by the chirality of the magnetic arrangement. 
{
More formally, this effect can be understood as an emergent gauge field generated by the noncollinear magnetic texture, which gives rise to a spin-dependent phase in the electron hopping processes. 
In this framework, the effective hopping acquires an additional term proportional to the vector spin chirality, $\boldsymbol{\kappa}_{ij} = \mathbf{m}_i \times \mathbf{m}_j$, leading to an effective Hamiltonian of the form
\begin{equation}
H_{\rm eff} \sim \sum_{\langle ij \rangle} c_i^\dagger \left[ \gamma + i \lambda\, (\boldsymbol{\kappa}_{ij} \cdot \boldsymbol{\sigma}) \right] c_j,
\end{equation}
which plays a role analogous to a spin--orbit interaction. 
}
As a consequence, the magnetic texture breaks the electronic spin rotational symmetry and allows for the emergence of an out-of-plane spin polarization, a hallmark of the spin Hall effect, 
{even in the absence of relativistic SOC~\cite{Katsura_PhysRevLett.95.057205, Hayami_PhysRevB.105.024413}}.
Here, our focus will be 
{on} the real-space manifestation of spin densities and currents in the kagome antiferromagnet.
{We note that the local out-of-plane spin density is not identical to the spin current, but reflects the spin character of nonequilibrium propagating states. Its relation to transport can be inferred from the corresponding current distribution.}

To investigate the real-space character of this unconventional, non-relativistic out-of-plane spin polarization, we consider a finite ribbon geometry that confines transport to a quasi-one-dimensional setup.
The geometry is illustrated in Fig.~\ref{fig1} for two different lattice terminations.
{
Figure~\ref{fig1}(a) shows the ribbon band structure together with the expectation value of the out-of-plane spin polarization.
Since the ribbon with equivalent upper and lower edge terminations preserves spatial inversion symmetry, the momentum-antisymmetric out-of-plane spin polarization,
$\langle \sigma_z(k)\rangle=-\langle \sigma_z(-k)\rangle$, is forbidden.
}
This corresponds to the case where the upper and lower edges of the ribbon are symmetric (see Fig.~\ref{fig1}).

{
When the two terminations are inequivalent, the top--bottom symmetry of the ribbon is broken, leading to the appearance of a momentum-antisymmetric spin splitting, even in the absence of relativistic SOC. 
A similar momentum-antisymmetric spin splitting has been predicted previously in kagome systems, although it typically requires asymmetric breathing of the lattice, which breaks the spatial inversion symmetry of the system~\cite{Hayami2020}. 
In the present case, however, the spin splitting is induced solely by the asymmetric ribbon confinement, which breaks the spatial inversion symmetry of the ribbon{, as the kagome plaquettes at the top edge become incomplete.} 
Consequently, the inversion-symmetry constraint prohibiting the momentum-antisymmetric spin splitting is removed, thereby allowing a finite momentum-antisymmetric spin splitting to emerge \cite{Hayami2020b}.
}

We further investigate the real-space transport in these two geometries. Figure~\ref{fig2} displays the spatial distribution of the out-of-plane spin density in the symmetric kagome ribbon for different transport energies. In panel (a), the ribbon supports $20$ open modes. Due to the symmetric confinement, the spin density alternates between up and down in a perfectly symmetric manner across the ribbon. In panel (b), only two electronic modes are available in the leads. In this regime, the spin density penetrates deeply into the bulk of the ribbon. Although the spin density is spatially extended, it averages to zero when integrated over the entire sample due to the preserved reflection symmetry. However, the opposite spin densities remain well separated in real space. 

{Further, we consider the complementary situation, where an asymmetric termination (depicted in Fig.~\ref{fig1}) is rather considered. The corresponding spin densities are shown in Fig. \ref{fig2}(c)-(d) for two different energies. In this configuration, the real space out-of-plane spin projection apparently remains symmetric.}
However, due to the asymmetry of the upper and lower edges, the spin response no longer averages to zero since the top--bottom symmetry is violated.

In this inequivalent-termination scenario, no exact reflection symmetry across the ribbon midline survives. In particular, the transverse spatial reflection $y \to -y$ does not map the lattice onto itself because the upper and lower edge terminations differ. Consequently, no combined magnetic mirror operation leaves the ribbon Hamiltonian invariant. The out-of-plane spin density is therefore not constrained to be either even or odd under reflection.

\begin{figure}
    \includegraphics[width=0.5\textwidth]{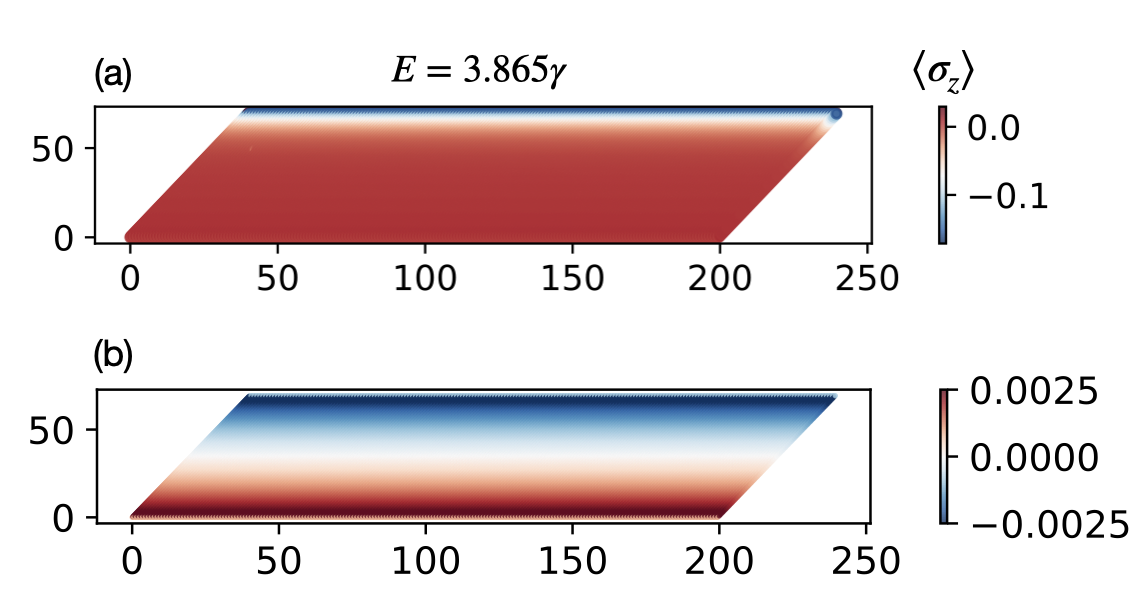}
    \caption{Spin resolved edge states for both {asymmetric} (a) and {symmetric} (b) ribbons. The energy is tuned such that only a single transport mode is enabled.}
    \label{fig3}
\end{figure}

In this situation, the two edges must instead be regarded as independent magnetic boundaries, each imposing its own local spin texture on the confined modes. As a result, isolated spin-polarized edge states can emerge, as illustrated in Fig.~\ref{fig3}(a).

\begin{figure}
    \includegraphics[width=0.5\textwidth]{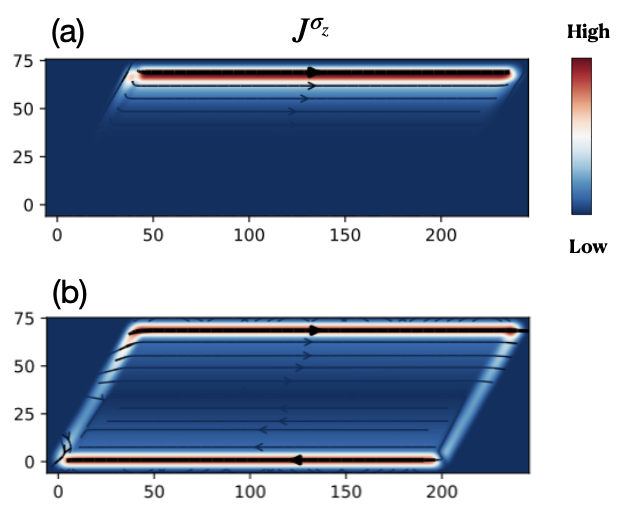}
    \caption{Spin current distribution in the ribbon for the edge states transport considered in Fig. \ref{fig3}. The edge spin currents within the asymmetric (symmetric) ribbons are shown in (a) and (b) respectively.}
    \label{fig4}
\end{figure}

By contrast, in the symmetric case{,} even a single transport channel must appear in pairs in order to satisfy the ribbon symmetry. 
The symmetric confinement forces the emergence of the polarized mode at both edges of the ribbon, as shown in Fig.~\ref{fig3}(b). 
{This distinction becomes clearer when compared with the corresponding current distribution shown below. While Fig.~3 visualizes the spatial structure of the out-of-plane spin polarization, the current density demonstrates that these spin-polarized regions are associated with propagating edge channels, indicating their relevance for spin transport.}
In Fig. \ref{fig4}{,} the current densities of the edge transport are also displayed. 

{
Although this behavior may appear unusual, it follows naturally from the coplanar $120^\circ$ magnetic order, whose moments lie in the $xy$ plane, together with the mirror symmetry with respect to the ribbon midline.}

\begin{figure}
    \includegraphics[width=0.5\textwidth]{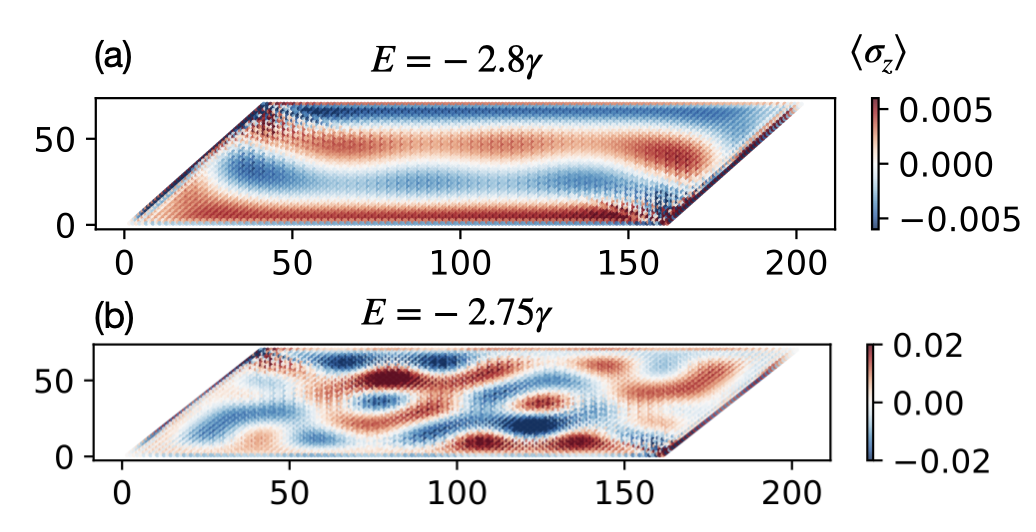}
    \caption{The real-space spin densities in the symmetric ribbon configuration with broken translational symmetry is shown for $E=-2.8\gamma$ ($E=-2.75\gamma$) in (a) and (b) respectively.}
    \label{fig5}
\end{figure}

{When the ribbon is attached to normal leads, the translational symmetry of the $120^\circ$ kagome magnetic texture along the ribbon direction is broken. 
The magnetic region becomes further {mirror symmetric with respect to the additional transverse mirror plane (perpendicular to the ribbon axis).
The combination of the two perpendicular mirror planes elevates the spatial symmetry of the system to an effective two-dimensional inversion symmetry around the center of the ribbon.
This leads to the emergence of an alternating spin polarization pattern around the center of the ribbon, as precisely illustrated in  Fig.~\ref{fig5}. This contrasts the previous scenario, where the spin polarization pattern is longitudinally invariant in the presence of magnetic leads, preserving the translational symmetry of the magnetic ribbon.}
}

A key distinction must be made between the local out-of-plane spin texture and the band-resolved spin polarization. In all ribbon terminations considered here, the confined states develop an alternating real-space $s_z$ pattern, indicating that the generation of local out-of-plane spin density is a generic consequence of confinement in the $120^\circ$ kagome phase. However, the visibility of this texture in the band structure depends strongly on the edge termination. Symmetric terminations tend to produce substantial cancellations between opposite edges and between different sublattice contributions, so that the net band-resolved expectation value of $\sigma_z$ remains weak even though the local $s_z$ texture is pronounced. By contrast, the inequivalent termination suppresses these cancellations, making the antisymmetric spin polarization directly visible in the ribbon band structure.

{ Note that a fixed width of 80 atoms has been considered throughout the present study. However, further simulations (not presented here) show that the bulk response remains qualitatively robust against changes in width, displaying characteristic oscillations in the spin density amplitudes. In contrast, the spin density of the edge states exhibits an exponential decay as a function of the ribbon width, consistent with the spatial confinement and decoupling of the boundaries.}

{
Finally,
it is worth comparing the present mechanism with the odd-parity spin polarization recently discussed for $p$-wave magnets \cite{Yu2025,Mitscherling2026}. 
In both cases, the momentum-antisymmetric spin splitting originates from inversion-symmetry breaking in the absence of relativistic spin--orbit coupling. 
The essential difference is that, in $p$-wave magnets, the inversion symmetry is broken spontaneously by the magnetic order itself, whereas in the present system it is broken extrinsically by the asymmetric ribbon termination. 
Thus, the odd-parity spin splitting reported here originates from boundary-induced symmetry breaking rather than from an intrinsic magnetic order parameter.
}

\section{Discussion}
The results presented above demonstrate that out-of-plane spin polarization can emerge in kagome antiferromagnets even in the absence of relativistic spin--orbit coupling. 
{This behavior} originates from {the noncollinear $120^\circ$ magnetic texture, which breaks spin rotational symmetry and gives rise to an effective spin-dependent electronic motion.} 
{In contrast to previous studies that primarily focused on momentum-space properties such as Berry curvature and spin currents, our results highlight how these effects manifest in real space under confinement. In particular, the ribbon geometry reveals an alternating spatial pattern of the out-of-plane spin polarization across the sample.
}

{The emergence and visibility of this spin texture are strongly governed by symmetry. In the symmetric ribbon, the presence of inversion symmetry enforces a cancellation of the out-of-plane spin polarization upon spatial averaging, even though a pronounced local texture exists. When this symmetry is broken by asymmetric edge termination, the cancellation is lifted, allowing a finite out-of-plane spin polarization to appear. This provides a direct symmetry-based understanding of the altermagnetic-like spin splitting observed in the system.
A distinctive feature of the confined system appears in the single-mode transport regime. In this case, a single propagating mode exhibits a spatial separation of spin polarization across the ribbon, with opposite spin character localized at opposite edges. This behavior differs from the conventional spin Hall effect, where separate spin channels propagate along different edges. Here, instead, a single mode carries a spatially varying spin texture.
}

{Importantly, the real-space spin polarization identified here is closely connected to transport properties. Although the local spin density is not identical to the spin current, its spatial distribution reflects the spin character of propagating states. Combined with the corresponding current distribution, this indicates that the observed spin textures are associated with conducting channels, particularly along the edges of the ribbon.
Overall, our results identify spatial confinement as a key ingredient that enables and reshapes spin polarization in noncollinear antiferromagnets without spin--orbit coupling. This establishes a direct link between magnetic texture, symmetry, and boundary geometry, providing a new perspective for engineering spin-dependent phenomena in nonrelativistic magnetic systems.
}

\section{Conclusion}
We have demonstrated that out-of-plane spin polarization can emerge in a kagome antiferromagnet with a coplanar $120^\circ$ magnetic texture in the absence of relativistic spin--orbit coupling. The effect originates from the breaking of spin rotational symmetry and the geometric phase acquired by electrons circulating in the chiral magnetic background, which effectively generates a spin-dependent hopping mechanism. In ribbon geometries, spatial confinement reveals a pronounced real-space alternating $s_z$ texture. While symmetric terminations lead to strong cancellations that suppress the net band-resolved spin polarization, inequivalent edge terminations lift these cancellations and produce an antisymmetric spin splitting reminiscent of altermagnetic states. Importantly, our results suggest that spin polarized edge states in the kagome antiferromagnet appears with a conceptual difference from edge transport in conventional spin Hall effect. Finally, our results suggest that asymmetric confinement provides a route to altermagnetic-like spin splitting in coplanar antiferromagnets.

\acknowledgments
{O. L thanks A. Manchon, U. Schwingenschlögl, H. Abdullah, A. Abbout, S. Murakami for useful discussions.}
{
S. H was supported by JSPS KAKENHI (Grants Nos. JP22H00101 and JP23H04869), 
and JST CREST (Grant No.~JPMJCR23O4), 
and JST FOREST (Grant No.~JPMJFR2366).
}


\bibliography{refs}
\end{document}